\documentclass[a4paper,11pt]{article}
\usepackage{jcappub}
\usepackage[utf8]{inputenc}
\usepackage{amsmath,mathtools}
\usepackage{tensor}
\usepackage{amssymb}
\usepackage{graphicx,epsfig,amssymb}
\usepackage{physics}
\usepackage{lipsum}  
\usepackage{bigints}
\usepackage{verbatim}
\usepackage{hyperref}
\usepackage{color}
\usepackage{multirow}
\usepackage{float}
\usepackage{array}
\newcolumntype{P}[1]{>{\centering\arraybackslash}p{#1}}

\usepackage{caption}
\usepackage{subcaption}
\usepackage[table,xcdraw]{xcolor}

\newcommand{\leri}[1]{\left(#1\right)}
\newcommand{\lerisq}[1]{\left[#1\right]}
\newcommand{\jor}{\frac{1}{2\kappa^2}\int d^4x\,\sqrt{-g}\;}
\newcommand{\ein}{\frac{1}{2\kappa^2}\int d^4x\, \sqrt{-q}\;}

\newcommand{\pal}{\mathcal{R}}

\usepackage[table,xcdraw]{xcolor}

\newcommand{\mary}[1]{{\textcolor{purple}{ [Mary: #1] } }}
\newcommand{\fla}[1]{{\textcolor{blue}{ [Fla: #1] } }}

\begin{document}
\title{Inflation in non-local hybrid metric-Palatini gravity}

\author[a]{Flavio Bombacigno,\note{Corresponding author.}}
\author[b]{Mariaveronica De Angelis,}
\author[b]{Carsten van de Bruck}
\author[b]{and William Giarè}

\affiliation[a]{Departament de F\'{i}sica Teòrica and IFIC, Centro Mixto Universitat de València - CSIC, Universitat de València, Burjassot 46100, València, Spain}
\affiliation[b]{School of Mathematical and Physical Sciences, University of Sheffield, Hounsfield Road, Sheffield S3 7RH, United Kingdom}

\emailAdd{flavio2.bombacigno@uv.es}
\emailAdd{mdeangelis1@sheffield.ac.uk}
\emailAdd{c.vandebruck@sheffield.ac.uk}
\emailAdd{w.giare@sheffield.ac.uk}

\abstract{
%
%
Within the framework of hybrid metric-Palatini gravity, we incorporate non-localities introduced via the inverse of the d'Alembert operators acting on the scalar curvature. We analyze the dynamical structure of the theory and, adopting a scalar-tensor perspective, assess the stability conditions to ensure the absence of ghost instabilities. Focusing on a special class of well-defined hybrid actions -- where local and non-local contributions are carried by distinct types of curvature -- we investigate the feasibility of inflation within the resulting Einstein-frame multi-field scenario. We examine how the non-minimal kinetic couplings between the fields, reflecting the non-local structure of the original frame, influence the number of e-folds and the field trajectories. To clarify the physical interpretation of our results, we draw analogies with benchmark single-field inflation scenarios that include spectator fields.
}
\makeatletter
\gdef\@fpheader{}
\makeatother

\maketitle
\section{Introduction}
General Relativity (GR) represents our current best theoretical framework for describing and understanding gravitational interactions. Since its introduction in 1915, it has consistently demonstrated its ability to explain well-known phenomena, such as Mercury's perihelion precession, which could not be accounted for within the context of Newtonian gravity. Moreover, it quickly became evident that GR predicted previously unforeseen phenomena, all of which have been corroborated by direct observations, including (but not limited to) the bending of light in gravitational fields~\cite{Walsh:1979nx}, the existence of black holes (BHs)~\cite{EventHorizonTelescope:2019dse}, and the propagation of gravitational waves (GWs)~\cite{LIGOScientific:2016aoc}.

Despite its remarkable success, GR faces several significant theoretical challenges. In strong-field regimes, such as near black holes and the Big Bang, the theory predicts singularities where physical quantities like curvature, temperature, or density diverge, causing physical laws (as we know them) to break down. Furthermore, all attempts to reconcile GR with quantum mechanics have so far been unsuccessful, limiting the theory's predictivity, especially in regimes where quantum gravitational effects -- such as spacetime fluctuations at the Planck scale -- are expected to play a crucial role. At these scales (possibly relevant for early Universe cosmology) GR struggles to provide a consistent framework, and quantum gravity effects are anticipated to dominate. Finally, GR is non-renormalizable because the gravitational coupling constant has dimensions, unlike the coupling constants in other fundamental forces, which are dimensionless. This leads to divergences in quantum field theory at high energies, meaning that quantum corrections produce infinite values that cannot be absorbed into a finite number of parameters. Consequently, GR becomes somewhat inconsistent in the high-energy limit, preventing it from being effectively integrated into quantum field theory frameworks.

To compound the challenges, when it comes to cosmological scales, GR and the Standard Model of particles (SM) on their own somehow fail to provide a coherent and satisfactory description of several observed phenomena. At the top of the list, we might mention the fact that the current accelerated expansion of the Universe sharply contrasts with GR’s predictions when only the matter content expected from the SM is considered. This problem led to re-introduce a cosmological constant term in the Einstein field equations (initially proposed by the same Einstein in 1917 to achieve a static Universe, reflecting the prevailing \textit{consensus} at the time that the Universe was unchanging and stationary), which is not free from conceptual problems~\cite{Sahni:1999gb, Carroll:2000fy, Peebles:2002gy, Padmanabhan:2002ji, Copeland:2006wr, Caldwell:2009ix, Li:2011sd, Martin:2012bt, Weinberg:1988cp, Krauss:1995yb, Weinberg:2000yb, Sahni:2002kh, Yokoyama:2003ii, Nobbenhuis:2004wn, Burgess:2013ara, Joyce:2014kja, Bull:2015stt, Wang:2016lxa, Brustein:1992nk, Witten:2000zk, Kachru:2003aw, Polchinski:2006gy, Danielsson:2018ztv, Zlatev:1998tr, Pavon:2005yx, Velten:2014nra} and seems challenged by recent observations~\cite{DESI:2024mwx,Giare:2024gpk,Giare:2024ocw}. Similarly, several independent observations, including the Cosmic Microwave Background (CMB) radiation and Large Scale Structure (LSS) of the Universe, contrast with what is expected in GR when only the matter content predicted by the SM is taken into account, providing indirect yet strong evidence for a missing mass in the universe commonly referred to as dark matter (DM). Last but not least, another de Sitter phase of accelerated expansion --  inflation~\cite{Guth:1980zm,Linde:1981mu,Albrecht:1982wi,Vilenkin:1983xq} -- appears necessary to set appropriate initial conditions and explain key features of the observable Universe, such as its remarkable flatness and the thermal equilibrium of the CMB radiation. The microphysical interpretation of these features remains unclear within our current understanding of fundamental interactions.\footnote{Many have argued that (part of) the tensions and anomalies emerging in recent years in cosmological and astrophysical data (most notably the Hubble tension~\cite{Verde:2019ivm,DiValentino:2020zio,DiValentino:2021izs,Perivolaropoulos:2021jda,Schoneberg:2021qvd,Shah:2021onj,Abdalla:2022yfr,DiValentino:2022fjm,Kamionkowski:2022pkx,Giare:2023xoc,Hu:2023jqc,Verde:2023lmm,DiValentino:2024yew} and the weak lensing discrepancy~\cite{DES:2021wwk,DiValentino:2020vvd,DiValentino:2018gcu,Kilo-DegreeSurvey:2023gfr,Troster:2019ean,Heymans:2020gsg,Dalal:2023olq,Chen:2024vvk,ACT:2024okh,DES:2024oud,Harnois-Deraps:2024ucb,Dvornik:2022xap}) may be traced back to limitations of GR on cosmological scales, hindering its ability to fully capture the dynamics of the Universe either at the background or perturbation level. Without claiming completeness, we refer readers interested in works exploring the impact of modified gravity theories on cosmological tensions to Refs.~\cite{Schiavone:2022wvq,Montani:2023xpd,Escamilla:2024xmz,Montani:2024xys,Banerjee:2022ynv,Petronikolou:2023cwu,Saridakis:2023pzo,Mandal:2023bzo,Bouche:2022qcv,Adil:2021zxp,Specogna:2023nkq,Specogna:2024euz,Ishak:2024jhs}}

Adopting a critical perspective, it is certainly worth pondering whether GR is the ultimate theory of gravitation, or rather the low-energy limit of a more fundamental theory capable of addressing both the theoretical challenges surrounding our current understanding of gravitational interactions and the missing pieces in the Universe's puzzle outlined above. In fact, this possibility has motivated a significant portion of the high-energy physics and gravity community to explore various extended theories of gravity beyond GR. 

Among the many scenarios explored, $f(R)$ theories of gravity stand out due to their conceptual simplicity, relevance, and flexibility. These theories have been extensively studied in various cosmological contexts~\cite{Carroll:2003wy}, demonstrating remarkable versatility in providing frameworks that can potentially account for a wide range of phenomena across energy scales differing by several orders of magnitude. At the heart of $f(R)$ gravity is the idea of replacing the Ricci scalar $R$ in the Einstein-Hilbert action of GR with a function $f(R)$, which introduces additional degrees of freedom. The central theme across different applications is the modified dynamical evolution of the gravitational degrees of freedom, which is driven by the extra scalar field embedded in the $f(R)$ formalism~\cite{Olmo:2005hc, Moretti:2019yhs}. While not exhaustive, it is worth noting that the inclusion of extra degrees of freedom in the effective gravitational action has proven to be a promising framework for addressing the dynamics of gravitating systems in the presence of DM~\cite{Sotiriou:2008rp, Nojiri:2017ncd, DeFelice:2010aj}. These theories have also been proposed as viable alternatives for triggering a phase of repulsive gravity responsible for the current accelerated expansion of the Universe within dark energy (DE) models, as well as for inducing nearly de Sitter dynamics in the early Universe in the context of inflationary cosmology. In fact, when it comes to inflation, it is no exaggeration to say that the most widely accepted models favoured by current observations -- most notably the one proposed by Alexei Starobinsky~\cite{Starobinsky:1980te} -- are grounded in $f(R)$ gravity (see, e.g., \cite{Starobinsky:1980te, Gottlober:1990um, Cognola:2007zu, Nojiri:2007cq, Artymowski:2014gea, Huang:2013hsb, Sebastiani:2015kfa, vandeBruck:2015xpa, Brooker:2016oqa} and references therein). 

That said, $f(R)$ gravity is not free from limitations. These theories are often inadequate in the weak-field regime and require a screening mechanism~\cite{Brookfield:2006mq, Faulkner:2006ub, Brax:2008hh, Burrage:2017qrf, Brax:2021wcv}. To address this and other limitations, alternative formulations to metric gravity have been proposed where the connection is treated as an independent variable, encapsulating local invariance under $GL(4)$ gauge transformations~\cite{Baldazzi:2021kaf}.  A notable example pertains to the Palatini gravity 
formalism, where the metric and the connection are varied independently in the action, resulting in distinct properties for the scalar field in the context of $f(R)$ gravity \cite{Olmo:2005hc,Olmo:2011uz}. While this approach has been able to account for the late-time cosmic acceleration models, 
significant limitations remains due to 
gradient instabilities in cosmological perturbations \cite{Koivisto:2005yc}. 
In the spirit of hitting two targets with the same arrow -- circumventing some of the shortcomings of both the Palatini and metric $f(R)$ approaches -- \textit{hybrid metric-Palatini gravity} has been devised in recent years~\cite{Harko:2011nh,Tamanini:2013ltp} (see also \cite{Karamitsos:2025ugq}) where both types of formulations are considered within a unique theoretical framework and  
a viable screening mechanism is already exhibited at the level of the Jordan frame \cite{Bombacigno:2019did}. The implications in the cosmology of the extended dynamical content have received some attention \cite{Capozziello:2012ny,Capozziello:2015lza,Rosa:2017jld}. 

An alternative approach, inspired by a quantum field theory perspective, to address the same limitations and challenges of GR involves considering non-local interaction terms in the Einstein-Hilbert action. The resulting non-local gravity theories, explored over the past decade, offer perspectives that are somewhat different from both $f(R)$ models and hybrid metric-Palatini gravity, see, e.g.,  Ref.~\cite{Capozziello:2021krv} and references therein for reviews. Long story short, depending on the type of non-locality implemented, these non-local theories can be distinguished in general terms in infinite derivative and integral kernel theories of gravity, based respectively on analytic transcendental functions of the d’Alembert operator $\Box$ or on the inverse operator $\Box^{-1}$. Infinite derivative models have demonstrated some interesting hints in addressing the issue of renormalizability, unitary and UV completion in gravitational theories beyond GR \cite{Briscese:2013lna,Koshelev:2016xqb}, as well as in preventing the appearance of singularities as byproduct of the non-local smearing mechanism \cite{Biswas:2011ar,Buoninfante:2018xiw}. Integral kernel theories were originally introduced in the seminal papers \cite{Deser:2007jk,Deffayet:2009ca,Deser:2013uya}, where it is was discussed the possibility of reproducing the late-time expansion of the Universe via the application of the $\Box^{-1}$ operator to the scalar curvature $R$. However, some criticalities of such an approach were outlined in Ref.~\cite{Belgacem:2018wtb}, and further extensions were proposed in Ref.~\cite{Deser:2019lmm} by authors in order to amend the original flaws.

Drawing inspiration from the distinct challenges and opportunities presented by non-local gravity, $f(R)$ models, and hybrid metric-Palatini gravity, this work takes a pioneering step forward by exploring the implications of incorporating non-localities into the hybrid metric-Palatini gravity framework, rigorously assessing the stability conditions of these combined theories and their applicability, particularly in the context of early Universe cosmology. We consider non-localities in the form of powers of the inverse of the d'Alembert operator, which we assume to act on both the type of curvatures, i.e. on the metric $R$ and the Palatini $\pal$ Ricci scalar. In doing so, we extend the analysis of the purely metric case of Ref.~\cite{DeFelice:2014kma} to embed also Palatini contributions, by following the hybrid formalism developed for $f(R,\pal)$ theories in Refs.\cite{Harko:2011nh,Tamanini:2013ltp,Bombacigno:2019did}. Furthermore, always in analogy with Ref.~\cite{DeFelice:2014kma}, we adopt the perspective of considering the non-local theory as equivalent to a \textit{local} scalar-tensor model, where non-localities are \textit{localized} via a suitable procedure relying on a set of auxiliary fields. That allows us to bypass the ambiguity of the retarded boundary conditions for the integral operator $\Box^{-1}$ (see discussion in Ref.~\cite{DeFelice:2014kma} and \cite{Carleo:2023xzj,Capozziello:2024bjb} for recent applications in gravitational wave phenomenology), as it was instead pursued in the original formulation\footnote{In this respect, it is important to highlight that the choice of a priori independent affine connection does not invalidate the assumptions of \cite{DeFelice:2014kma}. As discussed in \cite{Delhom:2020hkb}, the generalization of the D’Alembert operator to curved spacetimes does not introduce additional non-Riemannian couplings for $\Box$ acting on scalar fields. This implies that torsion and non-metricity do not enter the definition of the wave operator, which is ultimately determined only by the metric. For a detailed and instructive analysis on the role of non-Riemannian geometry in the generalization of quantum field theories from Minkowski to curved space-times see \cite{Delhom:2020hkb}.}\cite{Deser:2007jk,Deffayet:2009ca} (see also Ref.~\cite{Deser:2019lmm}). 
\\We thereby demonstrate that a na\"{i}ve extension of Ref.~\cite{DeFelice:2014kma} to the hybrid metric-Palatini case is not sufficient to remove the presence of ghost instabilities, whose number we show to depend on the sum of the highest powers of the $\Box^{-k}$ operators acting on the metric and the Palatini curvature. We discuss how this is an unavoidable property of every non-degenerate non-local $F(R,\pal, ..., \Box^{-m} R, ..., \Box^{-n}\pal )$ action, even when a purely Palatini approach is enforced, in contrast with the findings of standard $f(\pal)$ gravity where no additional degrees of freedom are excited. Following these considerations, we look then at specific configurations where degeneracy is explicitly violated, consisting in metric (Palatini) $f(R)$ ($f(\pal)$) models supplemented by Palatini (metric) non-local terms, where non-localities are linearly coupled to the curvature and introduce deformations to the Starobinsky-like potential, providing a novel path to test the robustness of the model.
We elucidate how such a hybridization mechanism is capable of restoring dynamical stability, and we derive a set of algebraic constraints assuring the absence of ghost modes in the corresponding three-dimensional scalar field space. 
In the second part of the work we address the feasibility of slow-roll inflation in such a theoretical framework, and as a preliminary step we assess the well-posedness of the first-order slow-roll parameter, ultimately resulting in additional constraints among the derivatives of the potential and the fields. We then numerically evaluate possible inflationary scenarios at the level of background evolution, analyzing how both the trajectories of the scalar fields over the potential and the e-fold number are affected by different choices of the initial non-localities. We focus in particular on quadratic metric $f(R)$ models accompanied by Palatini non-local terms, which turn out to be the only viable scenario for a finite slow-roll phase, being the quadratic Palatini $f(\pal)$ case with metric non-localities plagued by an infinite slow-rolling stage along one scalar field direction. Eventually, for every reliable scenario exhibiting the adequate number of e-folds, we numerically check \textit{a posteriori} the consistency of the no-ghost and slow-roll conditions along the dynamical evolution of the scalar fields.
\\ \indent The paper is organized as follows. In Sec.~\ref{sec: 2} we discuss the general framework, elucidating how ghost instabilities are introduced at the level of the scalar-tensor representation in the Einstein frame. In Sec.~\ref{sec: 3} we analyse two special configurations where local and non-local terms are carried by different types of curvature, which we proved to be dynamically stable. In Sec.~\ref{sec: 4} the general conditions for having a well-behaved slow-roll inflationary phase are set, and we numerically determine the e-fold number and the trajectories of the fields over the potential for different initial non-local terms, ultimately resulting in distinct non-diagonal kinetic terms and potential contributions in the Einstein frame. Finally, conclusions are drawn in Sec.~\ref{sec: 5}.
\\ \indent Conventions about the formalism adopted and the fundamental constants are established as follows. The gravitational coupling is set as $\kappa^2=8\pi G$, with $c=1$, and the spacetime signature is chosen mostly plus. The definition of the Riemann tensor we used is displayed by $\mathcal{R}\indices{^\rho_{\mu\sigma\nu}}=\partial_\sigma\Gamma\indices{^\rho_{\mu\nu}}-\partial_\nu\Gamma\indices{^\rho_{\mu\sigma}}+\Gamma\indices{^\rho_{\tau\sigma}}\Gamma\indices{^\tau_{\mu\nu}}-\Gamma\indices{^\rho_{\tau\nu}}\Gamma\indices{^\tau_{\mu\sigma}}$.

\section{Non-localities for hybrid metric-Palatini gravity}\label{sec: 2}
The starting point of our analysis is the non-local model discussed in Ref.~\cite{DeFelice:2014kma}, that here we extend along the lines of the hybrid framework introduced in Refs.~\cite{Tamanini:2013ltp,Bombacigno:2019did}. We consider the action
\begin{equation}
    S=\jor F(R,\pal, \Box^{-1} R, ..., \Box^{-m} R, \Box^{-1}\pal, ..., \Box^{-n}\pal ),
    \label{eq: general action}
\end{equation}
where $F$ is a function both of the Ricci curvature $R$, which is built from the Levi-Civita connection for the metric $g_{\mu\nu}$, and the Palatini Ricci curvature $\mathcal{R}$ depending instead on the independent connection $\Gamma\indices{^\rho_{\mu\nu}}$, i.e.
\begin{align}
    &R=g^{\mu\nu}R_{\mu\nu}(g)\\
    &\mathcal{R}=g^{\mu\nu}\mathcal{R}_{\mu\nu}(\Gamma).
\end{align}
In order to put the original action in a scalar-tensor form, we rewrite Eq.~\eqref{eq: general action} as
\begin{equation}
\begin{split}
       \jor \Biggl(F(R,\pal, \Vec{\alpha},\Vec{\beta})&-\lambda_1 (\Box \alpha_1-R)-\rho_1 (\Box \beta_1-\pal)\,+\Biggr.\\ &-\Biggr.\sum_{i=2}^m\lambda_i(\Box \alpha_i-\alpha_{i-1} ) - \sum_{j=2}^n\rho_j(\Box \beta_j-\beta_{j-1} )\Biggr),
\end{split}
\end{equation}
where we introduced the $n$-tuple and $m$-tuple of real scalar fields $\alpha_i,\beta_j$, denoted respectively by $\vec{\alpha}$ and $\vec{\beta}$, and the related Lagrange multipliers $\lambda_i,\rho_j$. Variation of Eq.~\eqref{eq: general action} with respect to $\lambda_i,\rho_j$ guarantees that the original formulation is consistently recovered, how it is showed by
\begin{align*}
    &\delta_{\lambda_1}S\rightarrow\alpha_1=\Box^{-1}R,\\&\delta_{\lambda_2}S\rightarrow\alpha_2=\Box^{-1}\alpha_1=\Box^{-2}R,\\
    &\dots\\
    &\delta_{\lambda_i}S\rightarrow\alpha_i=\Box^{-1}\alpha_{i-1}=\Box^{-i}R,
\end{align*}
and analogously for $\rho_j$ and the Palatini contribution. Then, the $F(R,\pal, \Vec{\alpha},\Vec{\beta})$ part can be further rearranged as
\begin{equation}
    \jor \leri{\frac{\partial F}{\partial  \chi}(R-\chi)+\frac{\partial F}{\partial \eta}(\pal - \eta)+F(\chi,\eta,\Vec{\alpha},\Vec{\beta})}~,
\end{equation}
so that by defining $F_{\chi}=\frac{\partial F}{\partial \chi}\equiv \phi$ and $F_{\eta}=\frac{\partial F}{\partial \eta}\equiv \xi$, we can rewrite the action as
\begin{equation}
\begin{split}
    \jor  \Biggl( (\phi+\lambda_1)R+&(\xi+\rho_1)\pal- W\leri{\phi,\xi,\Vec{\alpha},\Vec{\beta},\Vec{\lambda},\Vec{\rho}}+\Biggr.\\ &+\Biggl.\sum_{j=1}^m\leri{\nabla_\mu\alpha_j\nabla^\mu\lambda_j}+\sum_{j=1}^n\leri{\nabla_\mu\beta_j\nabla^\mu\rho_j} \Biggr)~,
\end{split}
\label{eq: scalar-tensor representation intermediate step}
\end{equation}
where we assumed $F_{\chi\chi}F_{\eta\eta}-F_{\chi\eta}^2\neq 0$ in order to have a well-defined inversion \begin{equation}
    \begin{cases}
        \chi &=f(\phi,\xi,\Vec{\alpha},\vec{\beta})\\
        \eta &=g(\phi,\xi,\Vec{\alpha},\vec{\beta})~.
    \end{cases}
\end{equation}
We also introduced the potential term
\begin{equation}
    W=\phi f(\phi,\xi,\Vec{\alpha},\Vec{\beta}) +\xi g(\phi,\xi,\Vec{\alpha},\Vec{\beta})-F(\phi,\xi,\Vec{\alpha},\Vec{\beta})-\sum_{j=1}^m \alpha_{j-1}\lambda_{j} - \sum_{j=1}^n \beta_{j-1}\rho_{j},
\end{equation}
provided the identification $\alpha_0,\beta_0=0$. Now, by varying Eq.~\eqref{eq: scalar-tensor representation intermediate step} with respect to the connection $\Gamma\indices{^\rho_{\mu\nu}}$, under the hypothesis of vanishing torsion ($\Gamma\indices{^\rho_{\mu\nu}}-\Gamma\indices{^\rho_{\nu\mu}}=0$) and metric compatibility ($\nabla_\rho g_{\mu\nu}=0$), it is possible to show that the equation for the connection takes the form
\begin{equation}
    \nabla_\rho \leri{\sqrt{-g}\,\Xi\, g^{\mu\nu}}=0,
\end{equation}
where $\Xi\equiv \xi+\rho_1$, which admits as solution the Levi-Civita connection for the conformal metric $h_{\mu\nu}=\Xi g_{\mu\nu}$, i.e.
\begin{equation}
\Gamma\indices{^\rho_{\mu\nu}}=\frac{1}{2}h^{\rho\sigma}\leri{\partial_\mu h_{\nu\sigma}+\partial_\nu h_{\mu\sigma}-\partial_\sigma h_{\mu\nu}}.
\end{equation}
In terms of the original metric $g_{\mu\nu}$ and the scalar field $\Xi$, this can be further rewritten as
\begin{equation}
    \Gamma\indices{^\rho_{\mu\nu}}=\frac{1}{2}g^{\rho\sigma}\leri{\partial_\mu g_{\nu\sigma}+\partial_\nu g_{\mu\sigma}-\partial_\sigma g_{\mu\nu}}+\frac{1}{2\Xi}\leri{\delta\indices{^\rho_\nu}\partial_\mu\Xi+\delta\indices{^\rho_\mu}\partial_\nu\Xi-g_{\mu\nu}\partial^\rho\Xi},
    \label{eq: sol connection expl}
\end{equation}
so that it is possible to express the Palatini curvature as
\begin{equation}
    \pal=R+\frac{3\nabla_\mu\Xi\nabla^\mu\Xi}{2\Xi^2}-\frac{\Box \Xi}{\Xi}.
    \label{eq: sol pal}
\end{equation}
We note that assuming from the very beginning a torsionless and metric-compatible connection is not mandatory, as a dynamical equivalent result can be obtained by solving the original equation of the connection for the different components of torsion and non-metricity. In this case, indeed, once the spurious degrees of freedom of the affine connection have been gauged out by virtue of the projective symmetry of the model, the non-Riemannian parts of the connection can be completely solved in terms of the derivatives of the scalar field $\Xi$, so that the final expression of the Palatini curvature is still displayed by \eqref{eq: sol pal} (see Refs.~\cite{Olmo:2005hc, Olmo:2011uz, Iosifidis:2018zjj} for technical details).
It is then possible to rearrange the action ~\eqref{eq: scalar-tensor representation intermediate step} into the form
\begin{equation}
\begin{split}
    \jor \Biggl( (\phi+\lambda_1+\Xi)R+&\frac{3\nabla_\mu\Xi\nabla^\mu\Xi}{2\Xi}- W\leri{\phi,\xi,\Vec{\alpha},\Vec{\beta},\Vec{\lambda},\Vec{\rho}} + \Biggr.\\ \Biggl. &+\sum_{j=1}^m\leri{\nabla_\mu\alpha_j\nabla^\mu\lambda_j}+\sum_{j=1}^n\leri{\nabla_\mu\beta_j\nabla^\mu\rho_j}\Biggr) ~.
\end{split}
\label{eq: scalar-tensor representation intermediate step 2}
\end{equation}
At this stage, we observe that is always possible to perform a linear field redefinition of the form
\begin{equation}
    \begin{cases}
        &\phi+\lambda_1+\Xi=\Phi\\
        &\lambda_i=a_{11}^{(i)}\psi_1^{(i)}+a_{12}^{(i)}\psi_2^{(i)}\\
        &\alpha_i=a_{21}^{(i)}\psi_1^{(i)}+a_{22}^{(i)}\psi_2^{(i)}\\
        &\rho_i=b_{11}^{(i)}\omega_1^{(i)}+b_{12}^{(i)}\omega_2^{(i)}\\
        &\beta_i=b_{21}^{(i)}\omega_1^{(i)}+b_{22}^{(i)}\omega_2^{(i)}~,\\
    \end{cases}
\end{equation}
which is well-defined as long as the Jacobian of the transformation is non trivial. Such a requirement leads in this case to the condition
\begin{equation}
    |J|=\prod_{i=1}^{m}\det A^{(i)}\prod_{j=1}^n\det B^{(j)}\neq 0,
\end{equation}
where we defined the submatrices
\begin{equation}
    A^{(i)}=\begin{pmatrix}
        a_{11}^{(i)} & a_{12}^{(i)} \\
        a_{21}^{(i)} & a_{22}^{(i)}
    \end{pmatrix},\qquad B^{(i)}=\begin{pmatrix}
        b_{11}^{(i)} & b_{12}^{(i)} \\
        b_{21}^{(i)} & b_{22}^{(i)}
    \end{pmatrix},
\end{equation}
and it is easy to see that $|J|\neq0$ simply amounts to require
\begin{equation}
   \det A^{(i)},\; \det B^{(j)}\neq 0\quad \forall \;i,j.
\end{equation}
Under these conditions, the action can be rearranged as
\begin{equation}
\begin{split}
    \jor \Biggl( \Phi R +&\frac{3(\nabla\Xi)^2}{2\Xi}-W(\Phi,\Xi,\Vec{\psi}_1,\Vec{\psi}_2,\Vec{\omega}_1,\Vec{\omega}_2)+\Biggr.\\ \Biggl.&+\sum_{i=1}^m K_{(i)}^{kl}\nabla_\mu\psi_k^{(i)}\nabla^\mu\psi_l^{(i)}+\sum_{j=1}^n H_{(j)}^{kl}\nabla_\mu\omega_k^{(j)}\nabla^\mu\omega_l^{(j)}\Biggr)~,
\end{split}
\end{equation}
with $k,l=1,2$ and
\begin{equation}
    K_{(i)}=\begin{pmatrix}
        a_{11}^{(i)}a_{21}^{(i)} & \frac{a_{11}^{(i)}a_{22}^{(i)}+a_{12}^{(i)}a_{21}^{(i)}}{2}\\
        \frac{a_{11}^{(i)}a_{22}^{(i)}+a_{12}^{(i)}a_{21}^{(i)}}{2}& a_{12}^{(i)}a_{22}^{(i)}
    \end{pmatrix},\qquad H_{(i)}=\begin{pmatrix}
        b_{11}^{(i)}b_{21}^{(i)} & \frac{b_{11}^{(i)}b_{22}^{(i)}+b_{12}^{(i)}b_{21}^{(i)}}{2}\\
        \frac{b_{11}^{(i)}b_{22}^{(i)}+b_{12}^{(i)}b_{21}^{(i)}}{2}& b_{12}^{(i)}b_{22}^{(i)}
    \end{pmatrix}.
\end{equation}
A simple realization for the matrices $A_{(i)},\,B_{(j)}$ is displayed by
\begin{equation}
    A_{(i)}=B_{(j)}=\begin{pmatrix}
        1 & \,\,\;1 \\
        1 & -1
    \end{pmatrix},
\end{equation}
which allows us to rewrite $K_{(i)}, H_{(j)}$ in the diagonal form
\begin{equation}
    K_{(i)}=H_{(j)}=\begin{pmatrix}
        1 & \,\,\; 0 \\
        0 & -1
    \end{pmatrix},\qquad \forall \;i,j
\end{equation}
so that we obtain
\begin{equation}
    S=\jor \leri{\Phi R +\frac{3(\nabla\Xi)^2}{2\Xi}+\leri{\mathbf{\Psi}+\mathbf{\Omega}}-W(\Phi,\Xi,\Vec{\psi}_1,\Vec{\psi}_2,\Vec{\omega}_1,\Vec{\omega}_2)},
\end{equation}
where we introduced the shortcut notation
\begin{align}
    &\mathbf{\Psi}\equiv\sum_{i=1}^m \leri{(\nabla\psi_1^{(i)})^2-(\nabla\psi_2^{(i)})^2}~,\\
    &\mathbf{\Omega}\equiv\sum_{j=1}^n \leri{(\nabla\omega_1^{(j)})^2-(\nabla\omega_2^{(j)})^2}~.
\end{align}
Eventually, we can rewrite the action in the Einstein frame defined by the conformal transformation $q_{\mu\nu}=\Phi g_{\mu\nu}$, which results in
\begin{equation}
   S_E= \ein \leri{R(q) -\frac{3(\nabla\Phi)^2}{2\Phi^2} +\frac{3(\nabla\Xi)^2}{2\Phi\Xi}+\frac{\leri{\mathbf{\Psi}+\mathbf{\Omega}}}{\Phi}-\frac{W(\Phi,\Xi,\Vec{\psi}_1,\Vec{\psi}_2,\Vec{\omega}_1,\Vec{\omega}_2)}{\Phi^2}}.
\end{equation}
It is clear, then, that by fixing the order $N=n+m$ of non-localities, the theory is always endowed with at least $N$ ghosts, irrespective of the sign of the field $\Phi$ and the form of the function $F$. Moreover, if we require that $\Phi>0$ for the conformal transformation to be well-defined, an additional ghost is present for $\Xi>0$, which agrees with the results of \cite{Bombacigno:2019did}. In particular, by selecting the $\Phi>0$ branch we can re-define the fields as
\begin{equation}
    \Phi=e^{\sqrt{\frac{2}{3}}\Phi_C},\qquad \Xi=\sigma_\xi\frac{\Xi_C^2}{6},
\end{equation}
where $\sigma_\Xi=\pm 1$ is the sign of the field $\Xi$,
and rewrite the action in its final form
\begin{equation}
    \ein \leri{R(q) -(\nabla\Phi_C)^2+\frac{\sigma_\xi(\nabla\Xi_C)^2+\mathbf{\Psi}+\mathbf{\Omega}}{ \,e^{\sqrt{\frac{2}{3}}\Phi_C}}-\frac{W(\Phi,\Xi,\Vec{\psi}_1,\Vec{\psi}_2,\Vec{\omega}_1,\Vec{\omega}_2)}{e^{2\sqrt{\frac{2}{3}}\Phi_C}}}.
\end{equation}
where all the kinetic terms have a canonical form up to a possible coupling with the field $\Phi_C$. In App.~\ref{app. a} we briefly discuss the peculiar case where only one type of curvature is considered in the initial action, showing how this does not alter the main conclusion about the structure and the number of ghost fields.

\section{Ghost free configurations
}\label{sec: 3}
In Ref.~\cite{DeFelice:2014kma} it was outlined that ghost instabilities can be removed by simply considering the linear coupling between the metric Ricci curvature and the nonlocal part of the action. Here we extend such a result, and we present two specific models where dynamics is stabilized by supplementing $f(R)$-like theories with nonlocal terms retaining the same kind of coupling of Ref.~\cite{DeFelice:2014kma}. In doing this, we follow the idea of Ref.~\cite{Harko:2011nh}, and we pursue \textit{hybridization} additively, by endowing the $f(R)$ term with a nonlocal part depending on the type of curvature not contained in the original $f$ function argument. That amounts to considering two possible configurations, displayed by the following Lagrangians:
\begin{equation}
    \mathcal{L}_1=f(\pal)+R\,G(\Box^{-1}R)-V(\Box^{-1}R),\qquad \mathcal{L}_2=f(R)+\pal\,G(\Box^{-1}\pal)-V(\Box^{-1}\pal).
    \label{eq: ghost free lagrangians}
\end{equation}
For the sake of completeness, we also included a function $V$ which does not affect the stability of the scalar modes (see discussion below) but plays the role of a potential term for the resulting scalar-tensor theory. We also note that configurations in Eq.~\eqref{eq: ghost free lagrangians} fall outside the discussion of Sec.~\ref{sec: 2}, in that condition $F_{RR}F_{\pal\pal}-F_{R\pal}^2\neq 0$ is now evaded. The first case we address is the Palatini $f(\pal)$ theory in the presence of metric nonlocalities as in $\mathcal{L}_1$, which by following the procedure illustrated in Sec.~\ref{sec: 2} can be recast in the following scalar-tensor form
\begin{align*}
    S_1=&\jor \leri{f(\pal)+R\,G(\Box^{-1}R)-V(\Box^{-1}R)}\\
    =&\jor \leri{\xi\pal-U(\xi)+(\lambda+G(\alpha))R+\nabla_\mu\alpha\nabla^\mu\lambda-V(\alpha)}\\
    =&\jor\leri{(\xi+\lambda+G(\alpha))R+\frac{3(\nabla\xi)^2}{2\xi}+\nabla_\mu\alpha\nabla^\mu\lambda-(V(\alpha)+U(\xi))}\\
    =&\jor \leri{\phi R+\frac{3(\nabla\xi)^2}{2\xi}+\nabla_\mu\alpha\nabla^\mu (\phi-\xi-G(\alpha))-W(\alpha,\xi)}~,
\end{align*}
where between the third and the fourth line we introduced the field $\phi\equiv \xi+\lambda+G(\alpha)$. Moving to the Einstein frame the action can be further rearranged as
\begin{equation}
    S_1=\ein\leri{\Tilde{R}-\frac{3(\nabla\phi)^2}{2\phi^2}+\frac{3(\nabla\xi)^2}{2\phi\xi}+\frac{\nabla_\mu\alpha\nabla^\mu (\phi-\xi-G(\alpha))}{\phi}-\frac{W(\alpha,\xi)}{\phi^2}},
    \label{eq: ein frame S1}
\end{equation}
with the kinetic matrix displayed by
\begin{equation}
    K_1=\frac{1}{\phi}\begin{pmatrix}
        \frac{3}{2\phi} & &0 & & -\frac{1}{2} \\ & & & &\\
        0 & & -\frac{3}{2\xi} & &\frac{1}{2} \\ & & & & \\
        -\frac{1}{2} & & \frac{1}{2}  & & G'(\alpha)
    \end{pmatrix}.
\end{equation}
We recall that the absence of ghosts is guaranteed if the kinetic matrix is positive definite, which usually is understood, for symmetric matrices, as the positiveness of its eigenvalues. Even if these can be found in principle by diagonalizing the kinetic matrix, due to the unknown dependence of the function $G$ on the scalar field $\alpha$, it is in general not possible to recover the functional relation between the old field base and the diagonal one, so that the original action cannot be explicitly rewritten in the latter. Therefore, in looking for equivalent definitions of positiveness which could ease the analysis, we resort to the so-called Sylvester's criterion, which for a symmetric real matrix allows us to consider the signs of the determinants of the upper left $k \times k$ submatrices $M_k$, with $1\le k \le n$, where $n$ is the dimension of the original matrix. By applying this to $K_1$ we then demand that the following conditions hold simultaneously:
\begin{align*}
    &||M_1||=\frac{3}{2\phi^2}>0, \\
    & ||M_2||=-\frac{9}{4\phi^3\xi}>0,\\
    & ||M_3||=\frac{3(\phi-\xi)-18 G'(\alpha)}{8\phi^4\xi}>0,
\end{align*}
and it is easy to check that the solution for the set of inequalities is given by $\phi>0,\xi<0,G'(\alpha)>\frac{\phi-\xi}{6}$ (we disregard the configuration with $\phi<0,\xi>0,G'(\alpha)<\frac{\phi-\xi}{6}$, since in this case the conformal transformation defining the Einstein frame is not well-behaved). Eventually, we note that some of the kinetic terms of $K_1$ can be put in their canonical form by redefining the fields as in Sec.~\ref{sec: 2}, i.e.
\begin{equation}
    \phi=e^{\sqrt{\frac{2}{3}}\Phi_c},\;\xi=-\frac{\Xi_c^2}{6},\;\Psi_c=\int \sqrt{G'(\alpha)}d\alpha,
    \label{fieldredefinition}
\end{equation}
with the definition of $\Psi_c$ which has to be understood for an assigned $G(\alpha)$ function. That allows us to rewrite Eq.~\eqref{eq: ein frame S1} as
\begin{equation}
\begin{split}
    S_1=\ein \Biggl( &\Tilde{R}-(\nabla\Phi_c)^2-\frac{(\nabla\Xi_c)^2+(\nabla\Psi_c)^2}{e^{\sqrt{\frac{2}{3}}\Phi_c}}+\Biggr.\\ \Biggl.&+\frac{d\alpha}{d\Psi_c}\nabla^\mu\Psi_c\leri{\sqrt{\frac{2}{3}}\nabla_\mu\Phi_c+\frac{\Xi_c}{3e^{\sqrt{\frac{2}{3}}\Phi_c}}\nabla_\mu\Xi_c}-\frac{W_1(\Psi_c,\Xi_c)}{e^{2\sqrt{\frac{2}{3}}\Phi_c}}\Biggr).
\end{split}
    \label{eq: ein fin S1}
\end{equation}
We stress that in this case the potential is separable in the fields $\Psi_c,\Xi_c$, up to a global factor depending solely on the field $\Phi_c$, i.e. 
\begin{equation}
    W_1(\Psi_c,\Xi_c)=V(\alpha(\Psi_c))+U\leri{-\frac{\Xi_c^2}{6}}.
    \label{eq: def W1}
\end{equation}
We anticipate that this peculiar configuration forbids a clear implementation of an inflationary scenario, in that it does not provide a clear mechanism for preventing the field $\Phi_c$ to slow roll indefinitely (as we discuss in more detail in Sec.~\ref{sec: 5}).
\\ For what concerns the model $\mathcal{L}_2$, it formally retains the same structure of $\mathcal{L}_1$, but with the two curvature interchanged, so that we deal with an initial metric $f(R)$ theory supplemented by Palatini nonlocalities. In this case the procedure of localization results in
\begin{align*}
    S_2=&\jor \leri{f(R)+\pal\,G(\Box^{-1}\pal)-V(\Box^{-1}\pal)}\\
    =&\jor \leri{\xi R-U(\xi)+(\rho+G(\beta))\pal+\nabla_\mu\beta\nabla^\mu\rho-V(\beta)}\\
    =&\jor\leri{(\xi+\rho+G(\beta))R+\frac{3(\nabla(\rho+G(\beta)))^2}{2(\rho+G(\beta))}+\nabla_\mu\beta\nabla^\mu\rho-(V(\beta)+U(\xi))}\\
    =&\jor \leri{\phi R+\frac{3(\nabla\psi)^2}{2\psi}+\nabla_\mu\beta\nabla^\mu (\psi-G(\beta))-W_2(\beta,\phi-\psi)},
\end{align*}
where now we redefined the fields as $\phi=\xi+\rho+G(\beta),\;\psi\equiv \rho+G(\beta)$, with $\beta$ kept untouched. In the Einstein frame, $S_2$ can be rearranged as
\begin{equation}
    S_2=\ein\leri{\Tilde{R}-\frac{3(\nabla\phi)^2}{2\phi^2}+\frac{3(\nabla\psi)^2}{2\phi\psi}+\frac{\nabla_\mu\beta\nabla^\mu (\psi-G(\beta))}{\phi}-\frac{W_2(\beta,\phi-\psi)}{\phi^2}}~,
\end{equation}
and the kinetic matrix is now displayed by
\begin{equation}
    K_2=\frac{1}{\phi}\begin{pmatrix}
        \frac{3}{2\phi} & &0 & & 0 \\ & & & &\\
        0 & & -\frac{3}{2\psi} & &-\frac{1}{2} \\ & & & & \\
        0 & & -\frac{1}{2}  & & G'(\beta)
    \end{pmatrix}.
\end{equation}
Following the discussion for $K_1$, we see that in this case the Sylvester's criterion gives us the conditions
\begin{align*}
    &||M_1||=\frac{3}{2\phi^2}>0,\\
    & ||M_2||=-\frac{9}{4\phi^3\psi}>0,\\
    & ||M_3||=-\frac{3(\psi+6G'(\beta))}{8\phi^4\psi}>0,
\end{align*}
whose only feasible solution is now $\phi>0,\psi<0,G'(\beta)>-\frac{\psi}{6}$. Then, by redefining the scalar fields as in the case of $S_1$, by simply trading the roles of $\xi,\alpha$ with $\psi,\beta$, the action $S_2$ takes the form
\begin{equation}
\begin{split}
    S_2=\ein \Biggl(\Tilde{R}-&(\nabla\Phi_c)^2-\frac{(\nabla\Xi_c)^2+(\nabla\Psi_c)^2}{e^{\sqrt{\frac{2}{3}}\Phi_c}}+\Biggr.\\ \Biggl. &+\frac{1}{3}\frac{d\beta}{d\Psi_c}\frac{\Xi_c}{e^{\sqrt{\frac{2}{3}}\Phi_c}}\nabla^\mu\Psi_c\nabla_\mu\Xi_c-\frac{W_2(\Phi_c,\Psi_c,\Xi_c)}{e^{2\sqrt{\frac{2}{3}}\Phi_c}}\Biggr).
\end{split}
    \label{eq: ein fin S2}
\end{equation}
In this case the potential displays the nice property of a $U$ component depending both on the $\Xi_c$ and the $\Phi_c$ fields,
\begin{equation}
    W_2(\Phi_c,\Psi_c,\Xi_c)=V(\beta(\Psi_c))+U\leri{e^{\sqrt{\frac{2}{3}}\Phi_c}+\frac{\Xi_c^2}{6}},
\end{equation}
which can potentially prevent the latter from an endless slow rolling phase. The functional dependence on $\Psi_c$ is still separable, and in general, as discussed in Sec.~\ref{sec: 5}, it can be conveniently neglected when interested into inflation. We conclude this section by noting that Eq.~\eqref{eq: ein fin S2} can be further manipulated into a diagonal form, but since it does not contribute to a significant improvement in the numerical analysis, we choose to report the corresponding expression in App.~\ref{app. b} in order to not overburden the mathematical exposition. Instead, we consider an application of the theory to inflationary cosmology in the very early universe. 

\section{Friedmann-Lema\^{i}tre-Robertson-Walker cosmology}\label{sec: 4}
In this section we perform a numerical analysis of the homogeneous and isotropic cosmological background described by the flat Friedmann-Lema\^{i}tre-Robertson-Walker line element
\begin{equation}
    ds^2 = - dt^2 + a(t)^2 (dx^2+dy^2+dz^2),
\end{equation}
with $a(t)$ denoting the scale factor. In order to present the equations of motion in a more compact way, it is useful to rewrite Eq.~\eqref{eq: ein fin S1} and Eq.~\eqref{eq: ein fin S2} in the concise form
\begin{equation}
\begin{split}
    S_i=\ein \Biggl( \Tilde{R}&-(\nabla\Phi)^2-\frac{(\nabla\Xi)^2+(\nabla\Psi)^2}{e^{\sqrt{\frac{2}{3}}\Phi}}+\Biggr.\\
    &\Biggl.+\frac{d\chi_i}{d\Psi}\nabla^\mu\Psi\leri{\sigma_i\sqrt{\frac{2}{3}}\nabla_\mu\Phi+\frac{\Xi_c}{3e^{\sqrt{\frac{2}{3}}\Phi}}\nabla_\mu\Xi}-Y_i\Biggr),
\end{split}
    \label{eq: ein fin gen}
\end{equation}
where $\chi_i=\alpha,\beta$ and $\sigma_i=1,0$ for $i=1,2$ respectively, and we dropped the $c$ subscript for the sake of clarity. We also introduced the generalized potential $Y_i=\frac{W_i}{e^{2\sqrt{\frac{2}{3}}\Phi}}$. Then, by varying Eq.~\eqref{eq: ein fin gen} with respect to $\Phi,\Xi$ and $\Psi$ we obtain 
\begin{align}
    \Box \Phi\;+ &\;\frac{e^{-\sqrt{\frac{2}{3}}\Phi}}{\sqrt{6}}\leri{(\nabla\Xi)^2+(\nabla\Psi)^2}-\frac{1}{2}\frac{\partial Y_i}{\partial \Phi}+\nonumber\\ &-\frac{\sigma_i}{\sqrt{6}}\nabla_\mu\leri{\frac{d\chi_i}{d\Psi}\nabla^\mu\Psi}-\frac{\Xi}{3\sqrt{6}}e^{-\sqrt{\frac{2}{3}}\Phi}\frac{d\chi_i}{d\Psi}\nabla_\mu\Psi\nabla^\mu\Xi=0,
    \label{eq: motion phi}\\
    \Box \Xi\;- & \; \sqrt{\frac{2}{3}}\nabla_\mu\Xi\nabla^\mu\Phi-\frac{e^{\sqrt{\frac{2}{3}}\Phi}}{2}\frac{\partial Y_i}{\partial \Xi}+\nonumber\\ &-\frac{\Xi}{6}\leri{\frac{d^2\chi_i}{d\Psi^2}(\nabla\Psi)^2-\sqrt{\frac{2}{3}}\frac{d\chi_i}{d\Psi}\nabla_\mu\Phi\nabla^\mu\Psi+\frac{d\chi_i}{d\Psi}\Box\Psi}=0,
    \label{eq: motion Xi}\\
    \Box \Psi\;- &\;\sqrt{\frac{2}{3}}\nabla_\mu\Psi\nabla^\mu\Phi-\frac{e^{\sqrt{\frac{2}{3}}\Phi}}{2}\frac{\partial Y_i}{\partial \Psi}+\nonumber\\ &-\frac{d\chi_i}{d\Psi}\leri{\frac{\sigma_i}{\sqrt{6}}e^{\sqrt{\frac{2}{3}}\Phi}\Box\Phi+\frac{\Xi}{6}\Box\Xi+\frac{(\nabla\Xi)^2}{6}-\frac{\Xi}{3\sqrt{6}}\nabla_\mu\Phi\nabla^\mu\Xi}=0.
     \label{eq: motion Psi}
\end{align}
Now, turning our attention to the equation for the metric field
\begin{equation}
\begin{split}
     &G_{\mu\nu}-\nabla_\mu\Phi\nabla_\nu\Phi-\frac{\nabla_\mu\Xi\nabla_\nu\Xi+\nabla_\mu\Psi\nabla_\nu\Psi}{e^{\sqrt{\frac{2}{3}}\Phi}}+\frac{d\chi_i}{d\Psi}\nabla_\mu\Psi\leri{\sigma_i\sqrt{\frac{2}{3}}\nabla_\nu\Phi+\frac{\Xi_c}{3e^{\sqrt{\frac{2}{3}}\Phi}}\nabla_\nu\Xi}+\\
     &+\frac{1}{2}g_{\mu\nu}\leri{(\nabla\Phi)^2+\frac{(\nabla\Xi)^2+(\nabla\Psi)^2}{e^{\sqrt{\frac{2}{3}}\Phi}}-\frac{d\chi_i}{d\Psi}\nabla^\rho\Psi\leri{\sigma_i\sqrt{\frac{2}{3}}\nabla_\rho\Phi+\frac{\Xi_c}{3e^{\sqrt{\frac{2}{3}}\Phi}}\nabla_\rho\Xi}+Y_i}=0,
\end{split}
\end{equation}
it is easy to derive from the $tt$--component the well-known Friedmann equation
\begin{equation}
    H^2=\frac{1}{6}\left( \dot{\Phi}^2+e^{-\sqrt{\frac{2}{3}}\Phi}(\dot{\Xi}^2+\dot{\Psi}^2) -\frac{d\chi_i}{d\Psi}\dot{\Psi}\leri{\sigma_i\sqrt{\frac{2}{3}}\dot{\Phi}+e^{-\sqrt{\frac{2}{3}}\Phi}\frac{\Xi}{3} \dot{\Xi}} +Y_i\right),
\end{equation}
whereas the $ij$--component results in 
\begin{equation}
    \dot{H}=-\frac{1}{2}\left( \dot{\Phi}^2+e^{-\sqrt{\frac{2}{3}}\Phi}(\dot{\Xi}^2+\dot{\Psi}^2) -\frac{d\chi_i}{d\Psi}\dot{\Psi}\leri{\sigma_i\sqrt{\frac{2}{3}}\dot{\Phi}+e^{-\sqrt{\frac{2}{3}}\Phi}\frac{\Xi}{3} \dot{\Xi}} \right),
\end{equation}
with $H \equiv \dot{a}/a$.
We require the leading order slow-roll parameter to be positive, i.e. $\epsilon_0\equiv -\frac{\Dot{H}}{H^2}>0$, usually corresponding in inflaton models with a single scalar field to neglect effective phantom dark energy scenarios ($w<-1$).
Moreover, we can recast the no-ghost condition in terms of the fields $\Phi,\Xi,\Psi$ as
\begin{equation}
    \frac{1}{6}\leri{\frac{d\chi_i}{d\Psi}}^2\leri{\sigma_ie^{\sqrt{\frac{2}{3}}\Phi}+\frac{\Xi^2}{6}}<1~.
    \label{eq:noghost}
\end{equation}
In the following subsections, we present and discuss the numerical results obtained by integrating the equations of motion outlined in this section, considering different case studies. We distinguish between two cases: \(V(\Box^{-1}\pal) = 0\) and \(V(\Box^{-1}\pal) \neq 0\). For each case, we examine the impact of different functional forms of the kinetic coupling. The algorithm used to integrate the equations of motion is based on previous work by some of us~\cite{DeAngelis:2023fdu,Giare:2023kiv,Cecchini:2024xoq,DeAngelis:2024jqc}, with modifications made to suit these particular cases.

\subsection{The case $V=0$}\label{subsec: 4.1}
\noindent In this section we specialize the analysis to models with $V(\Box^{-1}\pal)=0$. As discussed in Sec.~\ref{sec: 3}, such a condition does not affect the dynamical stability of the theory, but just deprives the global potential of the dependence on the field $\Psi$. Furthermore, we consider for the local $f(R)$ ($f(\pal)$) part a simple quadratic correction to the standard GR term, with hybrid non-localities given by the function $G$. That amounts to consider actions of the form
\begin{align}
    \mathcal{L}_1=a_1 \pal + b_1 \pal^2 + \,R\, G(\Box^{-1} R),\qquad 
    \mathcal{L}_2=a_2 R + b_2 R^2 +\,\pal\,G(\Box^{-1} \pal),
    \label{eq: lagr2}
\end{align}
which by following the procedure described in Sec.~\ref{sec: 3} result in the Einstein frame potential:
\begin{equation}
    Y_i(\Phi,\Xi,\Psi)=\frac{\leri{(1-\sigma_i )e^{\sqrt{\frac{2}{3}}\Phi}+(-1)^{\sigma_i}\frac{\Xi^2}{6}-a_i}^2}{4b_ie^{2\sqrt{\frac{2}{3}}\Phi}}.
    \label{eq: potential lin nonloc coupling0}
\end{equation}
The existence of a global minimum is a necessary property we demand in broad terms the potential to be endowed with, in order for the slow-rolling phase to terminate at some point in the trajectories' space for the scalar fields and the reheating stage to be possible to settle. By inspection of Eq.~\eqref{eq: potential lin nonloc coupling0}, we see then that for $\sigma_1=1$ the potential $Y_1$ is independent of the field $\Phi$, so that at least in the $\Phi$-direction the fields are expected to keep to slow roll, even if eventually they reach the minimum in the subspace spanned by $\Psi$ and $\Xi$. In this work, we are mainly interested in slow-rolling realizations of inflationary scenarios, and we leave for a subsequent analysis the discussion of alternative mechanisms in multi-field cosmology, e.g. ultra-slow roll, hybrid or rapid turn inflation. Our primary goal is indeed to demonstrate that a hybrid non-local extension of standard $f(R)$ gravity is theoretically compatible with an inflationary paradigm at least at the level of background dynamics, so that we look for the simplest realization of such a configuration, leaving aside for the moment the issues related to the role of perturbations. For all these reasons, we consider from now on only the case $\sigma_2=0$, where the potential $Y_2$ simplifies in
\begin{equation}
    Y_2(\Phi,\Xi)=\frac{\leri{e^{\sqrt{\frac{2}{3}}\Phi}+\frac{\Xi^2}{6}-a_2}^2}{4b_2e^{2\sqrt{\frac{2}{3}}\Phi}}.
    \label{eq: Y2}
\end{equation}
In general, we seek a well-shaped plateau potential with a minimum to enable inflation to occur by choosing, in the numerical implementations, suitable values for $a_2$ and $b_2$. However, the shape of the potential is not tightly constrained by these chosen values. Indeed, by requiring the potential to be positive we obtain the constraints $a_2>0$ and $b_2>0$, with $a_2$ determining the location of the minimum and $\frac{1}{b_2}$ the height of the potential.
For this configuration the no-ghost condition can be rearranged as
\begin{equation}
    0<\leri{\frac{d\chi_2}{d\Psi}\frac{\Xi}{6}}^2<1,
    \label{eq:noghost sigma2}
\end{equation}
which according to the form of the function $\chi_2(\Psi)$ determines the sub-region of plane $\{\Xi,\Psi\}$ where the motion of the scalar fields must be confined. We remark that for $\sigma_2=0$ the field $\Phi$ does not enter the former inequality and motion is not a priori constrained in the $\Phi$-direction. The explicit dependence of $\chi_2$ on $\Psi$ is established by the original choice of the non-local coupling $G$, but since the $\Psi$-factor in Eq.~\eqref{eq:noghost sigma2} is the same appearing in the equations of motion \eqref{eq: motion phi}-\eqref{eq: motion Psi}, it is more reasonable for the sake of numerical computation to deal directly with different possible choices of $K(\Psi)\equiv\frac{d\chi_2}{d\Psi}$. Specifically, we will discuss in the following the two configurations displayed by
\begin{equation}
    K(\Psi)=k \Psi^{1+n},\qquad K(\Psi)=k e^{n \Psi},
\end{equation}
that we dub respectively power-law and exponential kinetic coupling case, inasmuch as $K(\Psi)$ settles the non-diagonal kinetic terms in \eqref{eq: ein fin gen}. The original function $G$ can be obtained by reversing the definition of $\Psi$, i.e.
\begin{equation}
    G-G_0=\int \leri{\frac{d\Psi}{d\chi_2}}^2\,d\chi_2,
    \label{eq: rev int 1}
\end{equation}
where $\Psi=\Psi(\chi_2)$ is derived by the inversion of
\begin{equation}
    \chi_2 - \chi_{2,0}=\int K(\Psi)\,d\Psi.
    \label{eq: rev int 2}
\end{equation}
Some attention must be paid to the integration constants $G_0,\chi_{2,0},\Psi_0$ (with the latter stemming out from the r.h.s. of Eq.~\eqref{eq: rev int 2}), in that they can lead to some undesired features in the original action \eqref{eq: ghost free lagrangians}. As an example, we note that a non-vanishing $G_0$ introduces at the level of the action a linear term in the type of scalar curvature not contained in the original $f(R)$ function, thereby evading the hypotheses of Sec.~\ref{sec: 3}. An analogous effect is also expected to be triggered by some (non-trivial) combination of $\chi_{2,0}$ and $\Psi_0$, and this implies that the set of integration constants $\{G_0,\chi_{2,0},\Psi_0\}$ is in principle constrained and some mutual conditions must be implemented not to violate dynamical stability. This can be for instance appreciated by looking at the constant coupling case $K=k$, where we have
\begin{equation}
    G(\chi_2)=G_0+\frac{1}{k^2}\leri{\chi_2-\chi_{2,0}},
\end{equation}
and the condition $G_0=\frac{\chi_{2,0}}{k^2}$ must be accordingly enforced. Therefore, working directly with $K(\Psi)$ allows us to avoid such subtleties and to look for safe configurations at the level of the scalar-tensor reformulation.
\subsubsection{Case I: power-law kinetic coupling}\label{case I}
\begin{figure}[h!]
\centering    \includegraphics[width=1\linewidth]{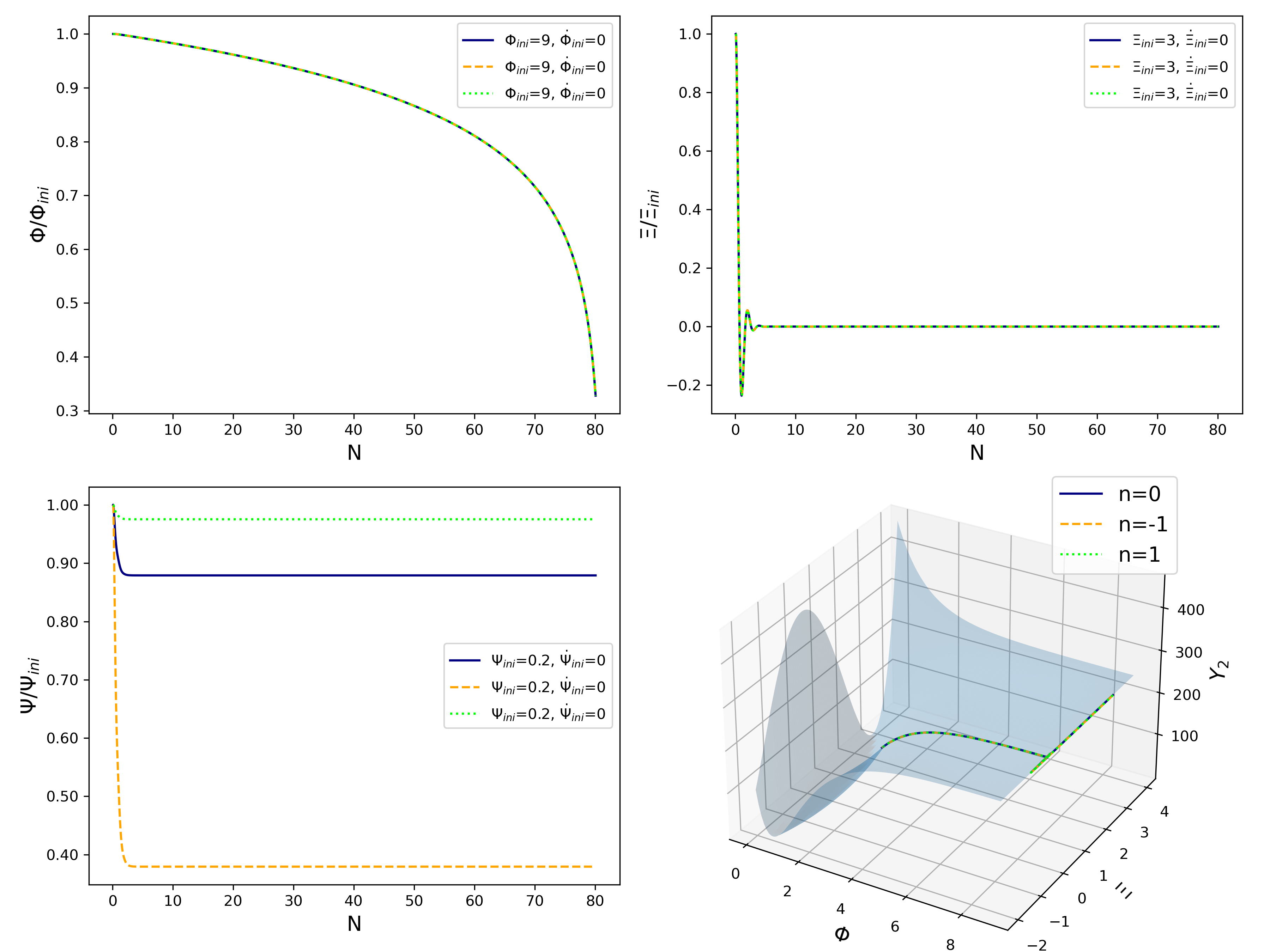}
    \caption{Power-law coupling case with $V = 0$: normalised fields trajectories and shaded potential profile $Y_2(\Phi, \Xi)$ with $a_2=2.3$, $b_2=0.001$, $k=0.1$ for $n=0$ (blue), $n=-1$ (orange) and $n=1$ (green).}
    \label{fig:eom1}
\end{figure}

\noindent The first case we analyse is the power-law kinetic coupling
\begin{equation}
K(\Psi)=k \Psi^{1+n},
\label{eq: power law kin coup}
\end{equation}
whose corresponding $G(\chi_2)$ function is displayed in App.~\ref{app: c}.
We begin by exploring the case where 
$n=-1$, which leads to a constant non-local term. This simplification reduces the complexity of the equations of motion, resulting in field trajectories that exhibit straightforward dependencies on the initial conditions. The kinetic terms introduce non-diagonal contributions in the Einstein frame, influencing the background evolution. From now on, we decide to consider the non-local contributions with $k \ll 1$ in order to study these terms acting as perturbations. In Fig.\ref{fig:eom1}, we integrate the whole system of equations \eqref{eq: motion phi}, \eqref{eq: motion Xi}, \eqref{eq: motion Psi} and it is carried up to the end of inflation which occurs for $\epsilon_0=1$. Numerical simulations indicate that inflationary scenarios are viable, with the fields gradually rolling along the potential (bottom right of Fig.\ref{fig:eom1}). However, careful selection of specific parameter values is essential to avoid ghost instabilities and to ensure a sufficient number of e-folds.
Indeed, we point out that in this case the no-ghost condition \eqref{eq:noghost sigma2} takes the form
\begin{equation}
    \left| k\; \Xi \;\Psi^{1+n}\right|<6.
\end{equation}
The central role of the inflaton field is underscored by the behavior of the $\Phi$-field, while the $\Xi$-field settles promptly to zero. Employing Eq.~\eqref{eq: Y2}, the potential lacks dependence on $\Psi$, and the coupling between $\Xi$ and $\Psi$ primarily arises from the non-local term. This coupling generates an initial kick to the $\Psi$ field, after which it experiences drag due to the expansion of the universe. It is worth noticing that when the dependence of $\Psi$ is turned on in Eq.~\eqref{eq: power law kin coup}, the non-local coupling becomes significant in the dynamics of the fields. Indeed, increasing the degree of Eq.~\eqref{eq: power law kin coup} the $\Psi$ field stabilises at higher values becoming almost completely frozen (see Fig.~\ref{fig:eom1}). This model can be treated in the same vein as single-field inflation with a spectator field,
given that its dynamical content is embedded in the $\Phi$ field only. Additionally, the above will be compared to $n=0$ considering $V\neq 0$ in the following section.

\subsubsection{Case II: exponential kinetic coupling}
\begin{figure}[h!]
\centering    \includegraphics[width=1\linewidth]{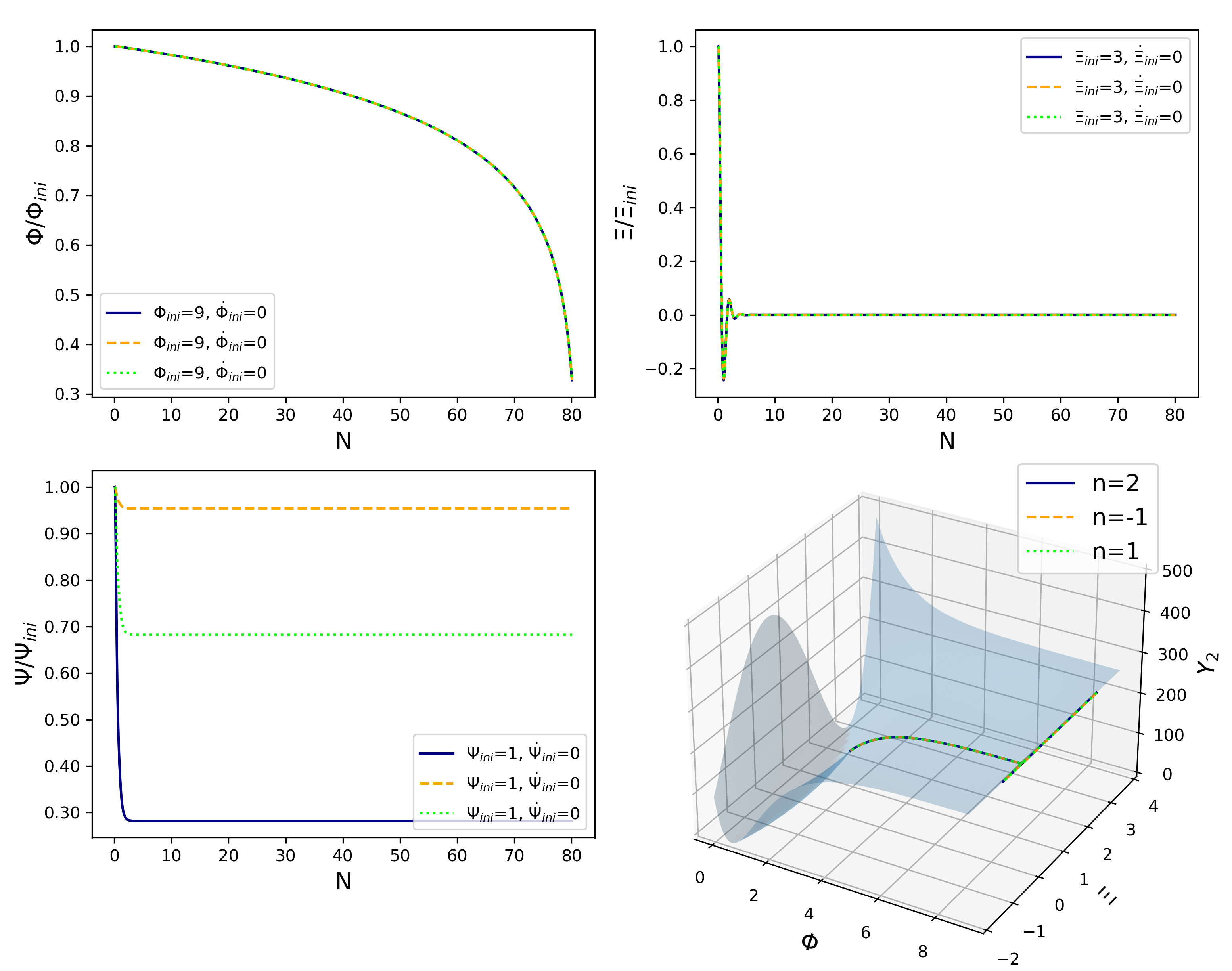}
    \caption{Exponential coupling case with $V = 0$: normalised fields trajectories and shaded potential profile $Y_2(\Phi, \Xi)$ with $a_2=2.3$, $b_2=0.001$, $k=0.1$ for $n=2$ (blue), $n=-1$ (orange), $n=1$ (green).}
    \label{fig:eom2}
\end{figure}

In this subsection we explore the following form 
\begin{equation}
    K(\Psi)=k e^{n\Psi},
    \label{eq: exp kin coup}
\end{equation}
with the corresponding $G(\chi_2)$ reported in App.~\ref{app: c}. Dynamical stability is satisfied for
\begin{equation}
    |k\, \Xi\, e^{n\Psi}|<6. 
\end{equation}
The exponential coupling again plays a crucial role in this context. As $n$ moves toward positive values, the interaction between $\Xi$ and $\Psi$ becomes more pronounced, allowing $\Psi$ to stabilise at lower values following an initial drop, Fig.\ref{fig:eom2}. Moreover, the field appears almost frozen reducing the dynamics to a single field scenario with a spectator field.

\subsection{The case $V\neq 0$}\label{subsec: 5.2}
In this section, we enlarge our analysis to encompass the case where a non-vanishing $V(\Box^{-1}\pal)$ function is considered in the initial action. This amounts to enriching the global potential $Y_2$ with a dependence on the field $\Psi$, obtained via the definition \eqref{eq: rev int 2}. In particular, following the discussion at the beginning of Sec.~\ref{sec: 4}, we select for $V$ a simple quadratic expression, i.e. $V(\chi_2)=V_0\chi_2^2$, where $V_0$ is constant so that the Einstein frame global potential is given by
\begin{equation}
    Y_2(\Phi,\Xi,\Psi)=\frac{V_0\chi_2^2(\Psi)+\leri{e^{\sqrt{\frac{2}{3}}\Phi}+\frac{\Xi^2}{6}-a_2}^2}{4b_2e^{2\sqrt{\frac{2}{3}}\Phi}}.
    \label{eq: potential lin nonloc coupling}
\end{equation}
\subsubsection{Case I: power-law kinetic coupling}
\begin{figure}[h!]
\centering    \includegraphics[width=1\linewidth]{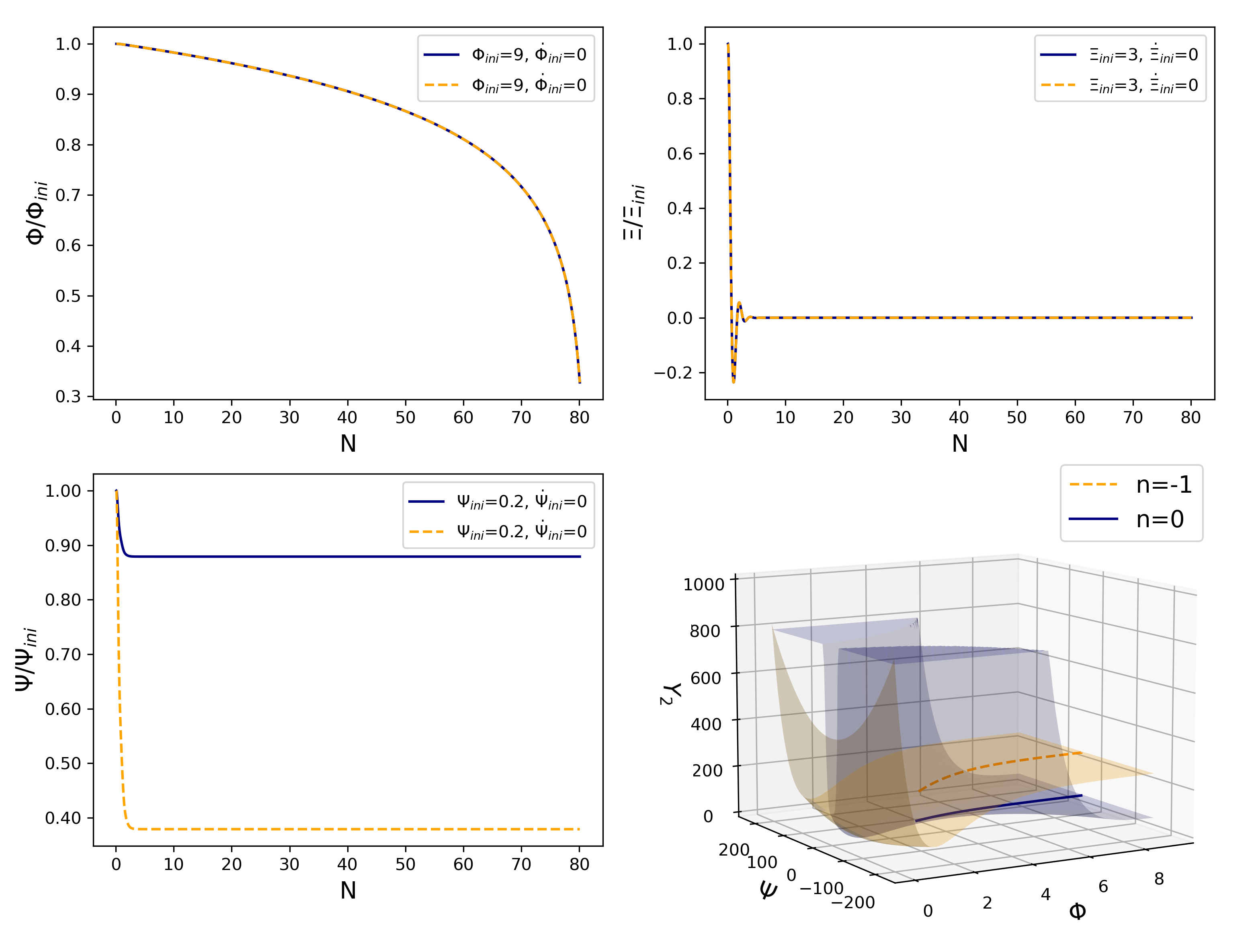}
    \caption{Power-law non-locality with $V \neq 0$: normalised fields trajectories and shaded potential profiles $Y_2(\Phi,0,\Psi)$ with $a_2=2.3$, $b_2=0.001$, $k=0.1$ for $n=0$ (blue) and $n=-1$ (orange).}
    \label{fig:eom1V}
\end{figure}

For $K(\Psi)$ as in Eq.~\eqref{eq: power law kin coup}, we have
\begin{equation}
    \chi_2=\frac{k}{n+2}\Psi^{n+2},
\end{equation}
where we set for the sake of simplicity the integration constants as $\chi_{2,0}=\frac{k\Psi_0^{n+2}}{n+2}$. For this choice, the potential term $V$ is displayed by
\begin{equation}
    V(\Psi)=\frac{k^2 V_0}{(n+2)^2}\Psi^{2(n+2)}.
\end{equation}
Focusing on $n=0$, where $K(\Psi) \approx \Psi$, there are no main differences from the case described in Subsec.\ \ref{case I}. 
Additionally, the condition $H^2/Y_{\Psi \Psi} \gg 1$  indicates that the field is light during inflation and hence it is Hubble damped and not driven to zero by the potential; instead, during inflation its kinetic energy drops and the field becomes constant. The change in the strength of the coupling also results in a change in potential amplitude (see Fig.~\ref{fig:eom1V}), which, in turn, would lead to an increase or decrease in the scalar spectral amplitude $A_s$ of primordial perturbations.

\subsubsection{Case II: exponential kinetic coupling}
For $K(\Psi)$ as in Eq.~\eqref{eq: exp kin coup}, we have
\begin{equation}
    \chi_2=\frac{k}{n}e^{n\Psi},
\end{equation}
where this time we set the integration constants as $\chi_{2,0}=\frac{k}{n}e^{n\Psi_0}$. In this case, the potential term $V$ is given by
\begin{equation}
    V(\Psi)= \frac{k^2 V_0 }{n^2}e^{2n\Psi}.
\end{equation}
\begin{figure}[h!]
\centering    \includegraphics[width=1\linewidth]{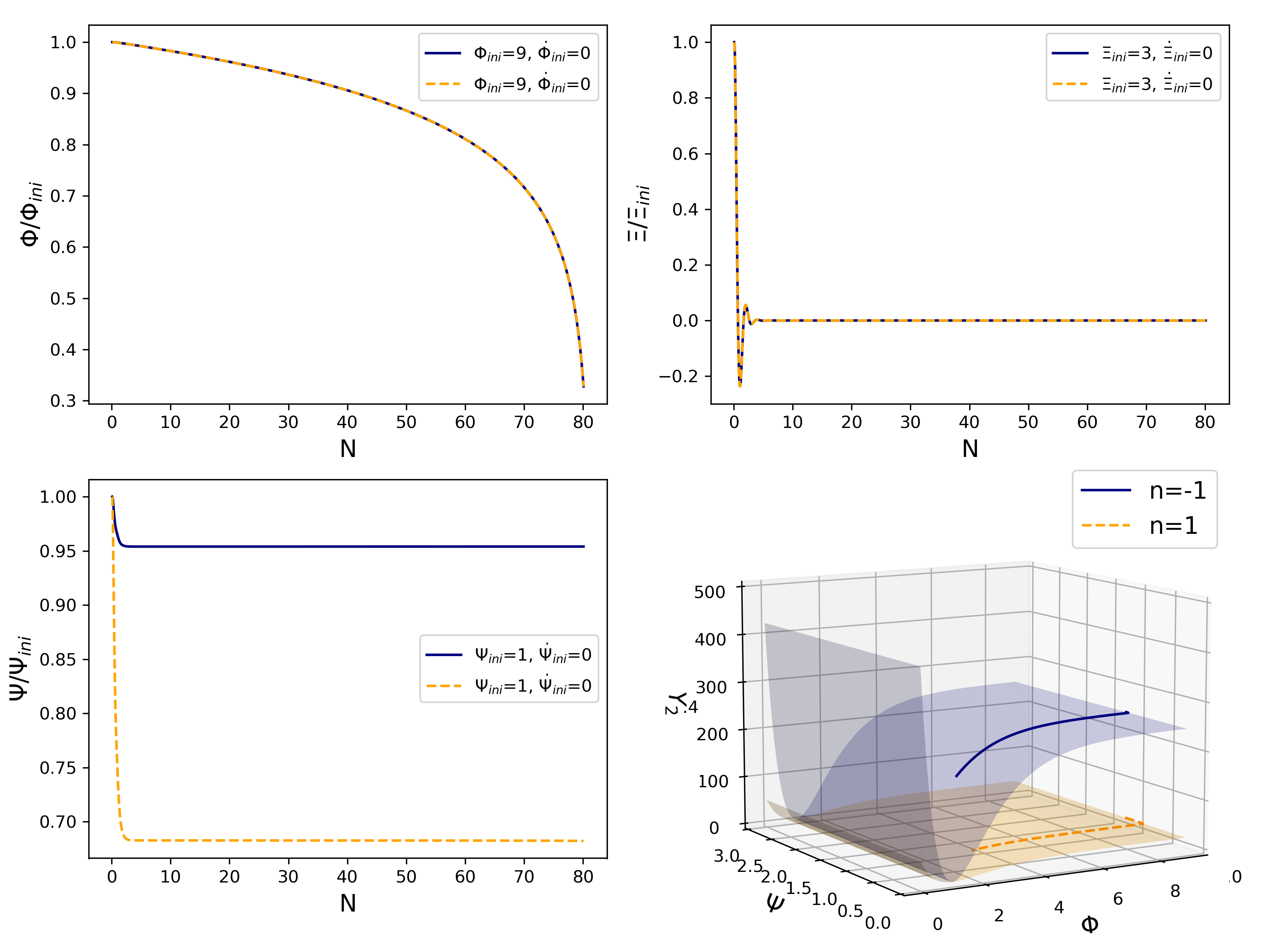}
    \caption{Exponential coupling with $V \neq 0$: normalised fields trajectories and shaded potential profiles $Y_2(\Phi,0,\Psi)$ with $a_2=2.3$, $b_2=0.001$, $k=0.1$ for $n=1$ (orange) and $n=-1$ (blue).}
    \label{fig:eom2V}
\end{figure}
The modification of the Starobinsky-like potential \eqref{eq: potential lin nonloc coupling} by $V(\Psi)$ for $\Phi \rightarrow 0$ allows inflation to happen, for a given shape of $V(\Psi)$, see Fig. \ref{fig:eom2V}. Also in this case, $H^2/Y_{\Psi \Psi} \gg 1$ is satisfied. The resulting field dynamics is very similar to the power-law case. 

\section{Conclusions}\label{sec: 5}
In this work, we explored non-local corrections in the form of powers of the inverse d'Alembert operator within the framework of hybrid metric-Palatini gravity. Specifically, we considered non-localities acting on both types of curvature: the metric Ricci scalar $R$ and the Palatini Ricci scalar $\pal$. We approached the non-local theory as equivalent to a local scalar-tensor model, where the non-local terms are localized through a procedure involving auxiliary fields, circumventing the ambiguity associated with imposing retarded boundary conditions for the integral operator $\Box^{-1}$. The main findings of our analysis can be summarized as follows: 
\begin{itemize}
    \item We rigorously demonstrated that non-degenerate hybrid metric-Palatini models supplemented by non-local terms built out of the inverse of the d'Alembert operators are generically plagued by ghost instabilities. These can be conveniently displayed by recasting the theory into a scalar-tensor framework, where the interplay between local and non-local terms is made explicit. We show that the number of ghost instabilities depends on the sum of the highest powers of the $\Box^{-k}$ operators acting on the metric and the Palatini curvature. We discuss how this is an unavoidable property of every non-degenerate non-local $F(R,\pal, ..., \Box^{-m} R, ..., \Box^{-n}\pal )$ action, even when a purely Palatini approach is enforced, in contrast with the findings of standard $f(\pal)$ gravity where no additional degrees of freedom are excited. This result extends the outcomes of Ref.~\cite{DeFelice:2014kma} for the purely metric case and is in agreement with the doubling of the degrees of freedom exhibited by generalized hybrid metric-Palatini gravity \cite{Tamanini:2013ltp,Bombacigno:2019did} with respect to the single curvature cases.
    
    \item We prove that Lagrangian densities characterized by the condition $f_{RR}f_{\pal\pal}-f_{R\pal}^2=0$ are sufficient for removing ghost instabilities, provided local and non-local terms are associated to distinct types of curvature. In particular, we established that simple modifications of well-known $f(R)$ ($f(\pal)$) theories with Palatini (metric) non-local terms, where curvature is linearly coupled to functions of the $\Box^{-1}$ operator, are fit for the purpose and can contain the additional dynamical content of the model to three scalar fields, where for comparison generalized hybrid models display two. 
    
\end{itemize}

Building on these considerations, in the second part of the work, we focused on a specific class of well-defined hybrid actions where local and non-local contributions were associated with distinct types of curvature. We examined configurations consisting of metric (Palatini) $f(R)$ ($f(\pal)$) models supplemented by Palatini (metric) non-local terms, where degeneracy was explicitly violated and non-localities were linearly coupled to the curvature. We investigated the applicability of these theories in the context of early Universe cosmology and studied the feasibility of inflation within the resulting Einstein-frame multi-field scenario. We analyzed both the role of non-localities and the influence of non-minimal kinetic couplings between the fields, reflecting the non-local structure of the original frame, on the background dynamics, the number of e-folds, and the field trajectories. As a preliminary step, we also assessed the well-posedness of the first-order slow-roll parameter, which ultimately resulted in additional constraints among the derivatives of the potential and the fields. Our analysis revealed that non-localities introduced deformations to the Starobinsky-like potential, providing a novel pathway to test the robustness of the model.

More quantitatively, we explored two key configurations motivated by the role played by non-local interaction terms in the resulting potential, considering two different scenarios: $V(\Box^{-1} \mathcal{R})=0$ and $V(\Box^{-1} \mathcal{R})\neq0$. In the first case, the dynamics are dominated by the coupling between the non-local terms and the scalar fields and we show that the behavior of the scalar fields does not depend critically on the choice of kinetic coupling $K(\Psi)$. Indeed, in both cases of power-law and exponential form, the $\Xi$ field promptly sets in the minimum of the potential while $\Psi$ freezes due to Hubble damping and $\Phi$ drives inflation, effectively reducing the system to a single-field scenario with a spectator field. Moreover, the stability of these trajectories hinges on satisfying the no-ghost condition. The inclusion of the potential term $V$ enriches the inflationary dynamics by allowing more intricate interactions between the fields. The quadratic form of $V$ introduces additional constraints that help terminate inflation. It is evident that the effective mass of the $\Psi$ field is light, implying that for general initial conditions it is not driven to zero during inflation. 

Overall, we demonstrated that non-local effects not only introduce an interplay between the dynamics of the fields by inducing deformations in the Starobinsky-like potential, but also have direct consequences on the background dynamics during the slow-roll phase of evolution. These effects leave characteristic footprints, offering a novel avenue for testing the robustness of the Starobinsky model and providing a significant first step toward characterizing a richer phenomenology that could help clarify the presence of potential non-localities in gravitational interactions at high energy scales characterizing the inflationary Universe. While a clear limitation of our analysis lies in focusing exclusively on the background-level effects of non-localities, our findings provide a solid foundation for further investigations into inflationary perturbations and their associated observational signatures within the framework of non-local hybrid metric-Palatini gravity.


\acknowledgments
The work of FB was supported by the postdoctoral grant CIAPOS/2021/169. MDA is supported by an EPSRC studentship. WG and CvdB are supported by the Lancaster–Sheffield Consortium for Fundamental Physics under STFC grant: ST/X000621/1. 

\appendix
\section{Ghosts for the purely metric and Palatini case}\label{app. a}
Following the procedure described in Sec.~\ref{sec: 2}, it is easy to check that when the initial function $F$ solely depends on one type of curvature, the final action in the Einstein frame can be recast as
\begin{equation}
    \ein \leri{R(q) -\delta_i\,(\nabla\Theta)^2+\frac{\mathbf{\Pi}}{ \,e^{\sqrt{\frac{2}{3}}\Theta}}-\frac{W(\Theta,\Vec{\pi}_1,\Vec{\pi}_2)}{e^{2\sqrt{\frac{2}{3}}\Theta}}},
    \label{eq: single case einstein frame}
\end{equation}
where $\delta_i=0,1$ for the Palatini and metric configuration, respectively, and we denoted with $\Theta$ and $\Pi$ the generic field representation associated to each case. We observe that for $\delta_i=1$ the results of \cite{DeFelice:2014kma} are reproduced, i.e. $N$ ghost fields appear, while selecting $\delta_i=0$ deprives $\Theta$ of proper dynamics, and in this case, it can be completely expressed in terms of the other fields, as in ordinary Palatini $f(\pal)$ theories. Indeed, variation of \eqref{eq: single case einstein frame} with respect to $\Theta$ results in the equation
\begin{equation}
    2W-\frac{\partial W}{\partial \Theta}=e^{\sqrt{\frac{2}{3}}\Theta}\mathbf{\Pi},
\end{equation}
which once $W$ is assigned, can be in principle solved for $\Theta=\Theta(\Vec{\pi}_1,\Vec{\pi}_2)$. We remark that also in this case $N$ ghost show up in $\mathbf{\Pi}$, so that the theory is still dynamically unstable.

\section{Diagonalization for $\sigma_2=0$}\label{app. b}
In this appendix we report the diagonal form for \eqref{eq: ein fin S2}, which is obtained by the field redefinition $\Psi=\eta+\omega$, $\Xi=\eta-\omega$, leading to:
\begin{equation}
\begin{split}
    S_2= \ein \Biggl( R-(\nabla\Phi)^2&-\frac{K_-(\eta,\omega)(\nabla\eta)^2+K_+(\eta,\omega)(\nabla\omega)^2}{e^{\sqrt{\frac{2}{3}}\Phi}}+\Biggr.
    \\ \Biggl.&-\frac{V(\beta(\eta+\omega))+U\leri{e^{\sqrt{\frac{2}{3}}\Phi}+\frac{(\eta-\omega)^2}{6}}}{e^{2\sqrt{\frac{2}{3}}\Phi}}\Biggr),
\end{split}
\end{equation}
where we introduced the kinetic coefficients
\begin{equation}
    K_{\mp}(\eta,\omega)\equiv 2 \mp \frac{\eta-\omega}{3}\left.\frac{d\beta}{d\Psi}\right|_{\Psi=\eta+\omega}.
\end{equation}
\section{Expressions of the $G$ function}\label{app: c}
Following the procedure outlined in Sec.~\ref{sec: 5}, the non-local $G$ functions corresponding to a power-law in the Einstein frame is given by:
\begin{align}
    &G(\chi_2)=G_0+\frac{1}{n k}\lerisq{\Psi_0^{-n}-\leri{\Psi_0^{n+2}+\frac{n+2}{k}\leri{\chi_2-\chi_{2,0}}}^{-\frac{n}{n+2}}},\qquad &&n\neq 2
    \label{eq: function G power law n not 2}\\
&G(\chi_2)=G_0+\frac{\Psi_0^2}{2k}\leri{e^{\frac{2(\chi_2-\chi_{2,0})}{k}}-1},\qquad &&n= 2 \label{eq: function G power law n 2}
\end{align}
while an exponential coupling results in 
\begin{equation}
    G(\chi_2)=G_0+\frac{\chi_2-\chi_{2,0}}{ke^{n\Psi_0}\leri{ke^{n\Psi_0}+n\leri{\chi_2-\chi_{2,0}}}}~.
    \label{eq: function G exp}
\end{equation}
\bibliographystyle{JHEP}
\bibliography{bib}
\end{document}